# Carrier Multiplication in Nanocrystals via Photostimulated Generation of Biexcitons from Vacuum

Valery I. Rupasov*

*ALTAIR Center LLC, Shrewsbury, MA 01545, USA*

*Landau Institute for Theoretical Physics, Moscow, Russia*

Victor I. Klimov†

*Chemistry Division, Los Alamos National Laboratory, Los Alamos, New Mexico 87545, USA*

We propose a novel mechanism for photogeneration of multiexcitons by single photons (carrier multiplication) in semiconductor nanocrystals. In this mechanism, the Coulomb interaction between two valence-band electrons involving their transfer to the conduction band creates a virtual biexciton from vacuum that is then converted into a real biexciton by photon absorption on an intraband optical transition. This mechanism is inactive in bulk semiconductors as momentum conservation suppresses intraband absorption. However, it becomes highly efficient in zero-dimensional nanocrystals and can provide a significant contribution to carrier multiplication in these materials.

Carrier multiplication (CM) is a process, in which absorption of a single photon produces not just one but multiple electron-hole (e-h) pairs (excitons). CM can benefit a number of technologies, especially, photovoltaics and photocatalysis. In bulk materials, CM has been typically explained by impact ionization [1]. In this process, a conduction-band electron or a valence-band hole of sufficiently high energy interacts with a valence-band electron promoting it across the energy gap ($E_g$). An important characteristic of CM is the e-h pair creation energy ($\varepsilon$) defined as the energy lost by the ionizing particle in a single e-h pair creation event. On the basis of energy conservation, the minimal value of $\varepsilon$ is $E_g$. However, because of restrictions imposed by momentum conservation and significant phonon losses, the values of $\varepsilon$ measured for bulk semiconductors significantly exceed this energy-conservation-defined limit. In particular, $\varepsilon$ is approximately $3E_g$ for wide-gap semiconductors and even greater for narrow-gap materials [2, 3].

Recent experimental studies of zero-dimensional (0D) semiconductor nanocrystals (NCs) [4-6] indicate that in these structures $\varepsilon$ approaches its ultimate limit of $E_g$, which results in extremely high CM efficiencies [6]. The mechanism for this enhancement is not well understood. For example, Zunger and co-workers adopted a traditional impact-ionization model for treating CM in NCs [7, 8]. On the other hand, recent calculations by Allan and Delerue [9] indicate that impact ionization is not enhanced by quantum confinement suggesting the existence of either additional or alternative mechanisms for CM in NCs.

Two such models have been discussed by Efros, Nozik and coworkers [5, 10] and Schaller, Agranovich, and Klimov [11]. Their common motif is that CM occurs via direct Coulomb coupling of a single-exciton to a biexciton state (matrix element, $U_{II}$, is the same as in the impact-ionization models). In Ref. [11], this process was treated using second-order perturbation theory as a transition via intermediate virtual single-exciton states, while Ref. [10]

utilized a time-dependent density-matrix approach. In the limit of weak Coulomb coupling ($U_{II} < \Gamma_{xx}$; $\Gamma_{xx}$ is the biexciton dephasing rate) both models lead to a similar result for the ratio of the biexciton ($N_{xx}$) to the single-exciton ($N_x$) populations. Specifically, in the case of a single intermediate exciton state (dephasing rate $\Gamma_x$) coupled to a single biexciton state, $N_{xx}/N_x = |U_{II}|^2/(\Gamma_x\Gamma_{xx})$. This expression shows that $N_{xx}/N_x$ can be greater than unity only if $U_{II} > \Gamma_x$, which was regarded as a necessary condition for efficient CM in Ref. [10]. However, as was pointed out in Ref. [11], the available experimental data indicate that Coulomb coupling in NCs is likely smaller than $\Gamma_x$. Therefore, it was suggested that an important factor contributing to the high efficiency of CM in NCs is a large density of biexciton states ($g_{xx}$), which can greatly exceed that of single-exciton states ($g_x$) of similar energies. Recent pseudo-potential calculations indeed indicate fast increase in the $g_{xx}$-to-$g_x$ ratio with increasing energy above the lowest biexciton resonance [8].

In this Letter, we use second-order perturbation theory to analyze a new aspect of the direct biexciton photogeneration model associated with the possibility of two different time orderings of the interaction terms responsible for photoexcitation and the Coulomb interaction. One time ordering, considered in Ref. [11], involves first *interband* optical excitation of an intermediate exciton state (virtual exciton), which then is converted into a biexciton via the impact-ionization term, $H_{II}$, of the Coulomb interaction (Fig. 1; left of the thick gray arrow). However, another possibility is generation of an intermediate biexciton state (virtual biexciton) from vacuum via the Coulomb interaction component, which involves transfer of two electrons from the valence to the conduction band, $H_{xx}$, (Fig. 1; right of the thick gray arrow); this intermediate state is then converted into a final biexciton state via an *intraband* optical transition initiated by the absorbed photon. The time ordering in this case is similar to that in one of the two channels for the Raman process, in which emission of a Stokes photon precedes absorption of an incident photon [12]. Furthermore, as in the Raman process, the total probability of biexciton photogeneration is determined by contributions from both virtual-exciton and virtual-biexciton channels. Using density-of-states arguments and taking into account the large strength of NC intraband transitions, we show that the CM rate for the mechanism involving intermediate biexciton states can be comparable to or higher than the single-exciton generation rate even for weak Coulomb coupling of ~1 meV. An important feature of this process is that it is inactive in bulk materials because of translational momentum conservation, which suppresses intraband optical transitions. On the other hand, because of relaxation of momentum conservation, the proposed mechanism can become highly efficient in 0D nanostructures. The existence of this additional CM channel, which operates in parallel with impact-ionization-like processes [11] (or possibly even dominates them), may explain unusually high efficiencies of CM in NCs.

Available experimental data indicate that CM in NCs does not follow the trends previously established for bulk semiconductors, which might be considered as a sign that this process occurs via different pathways in these two classes of materials. For example, Fig. 2(a) compares quantum efficiencies ($Q$) for conversion of photons into e-h pairs measured as a function of photon energy ($\hbar\omega$) for bulk PbS [13] as well as PbS NCs and PbSe NCs [6]. This comparison indicates a considerably lower CM threshold in NCs (~$3E_g$) than in the bulk (~$5E_g$). The other difference is in the slope of the $Q$-vs-$\hbar\omega$ dependence, which provides a direct measure of the e-h pair creation energy: $\varepsilon = \mathrm{d}(\hbar\omega)/\mathrm{d}Q$. Based on the data in Fig. 2(a), $\varepsilon = 4.5E_g$ in bulk PbS, while it is ca. $E_g$ in NCs.

The e-h creation energy is normally calculated as a sum of $E_g$, the average kinetic energy of carriers following the impact-ionization event ($<E_K>$), and the energy lost to phonons ($<E_{ph}>$) [2, 3]. Since $\varepsilon$ measured for NCs is close to $E_g$, it implies that both $<E_K>$ and $<E_{ph}>$ are approaching zero. The fact that $<E_K> = 0$ can be attributed to relaxation of momentum conservation, while the zero value of $<E_{ph}>$ indicates that CM in NCs occurs without measurable phonon losses. The latter is an important conclusion, which allows one to treat CM not as a competition between impact ionization and phonon-assisted relaxation (traditional approach used for bulk semiconductors) but as *direct* photogeneration of biexcitons via intermediate either single-exciton [11] or biexciton states (Fig. 1).

The first of these two pathways is active in both bulk and NC materials, while the second is unique to 0D materials because it involves the final step mediated by intraband absorption, which is suppressed in bulk semiconductors. Therefore, this latter process could, in principle, explain the enhancement of CM efficiencies in NCs. The corresponding term of the Coulomb interaction describes scattering between two valence-band electrons accompanied by their transfer to the conduction and it can be presented as as:

$$H_{xx} = \frac{1}{4} \sum_{a_1 a_2 b_1 b_2} \left[ U_{b_1 b_2}^{a_1 a_2} c_{a_1}^\dagger c_{a_2}^\dagger d_{b_1} d_{b_2} + h.c. \right], \qquad (1)$$

$$U_{b_1 b_2}^{a_1 a_2} = \int d\mathbf{r}_1 d\mathbf{r}_2 \Psi_{a_1}^*(\mathbf{r}_1) \Psi_{a_2}^*(\mathbf{r}_2) U_C(|\mathbf{r}_1 - \mathbf{r}_2|) \left[ \Psi_{b_1}(\mathbf{r}_2) \Psi_{b_2}(\mathbf{r}_1) - \Psi_{b_2}(\mathbf{r}_2) \Psi_{b_1}(\mathbf{r}_1) \right], \quad (2)$$

where $U_C$ is the Coulomb interaction energy, $c^\dagger$ ($c$) and $d^\dagger$ ($d$) are electron creation (annihilation) operators for the conduction and the valence bands, respectively, and $\Psi$ is the electron wave function for the conduction (index "a") or the valence (index "b") bands. Acting on the NC vacuum state $|0\rangle$, operator $H_{xx}$ creates two electrons in the conduction band and two holes in the valence band, i. e. a biexciton. Since energy is not conserved in this process, it does not contribute to transition rates calculated within first-order perturbation theory. However, its contribution is nonzero in second-order perturbation theory. In this case, the intermediate, virtual biexciton state, $\left| x_1' x_2' \right\rangle$, generated from vacuum by the operator $H_{xx}$ is converted into a real biexciton, $\left| x_1 x_2 \right\rangle$, by the intraband optical transition (Fig. 1; right of the thick gray arrow). This two-step process produces a final state, which satisfies energy conservation and the corresponding rate can be calculated as

$$W_2 = \frac{2\pi}{\hbar} \sum_{x_1 x_2} \left| \sum_{x_1' x_2'} \frac{\left\langle x_1 x_2 \left| H_{ev}^{\mathrm{intra}} \right| x_1' x_2' \right\rangle \left\langle x_1' x_2' \left| H_{xx} \right| 0 \right\rangle}{E_{x_1' x_2'}} \right|^2 \delta(\hbar\omega - E_{x_1 x_2}), \qquad (3)$$

where $\hbar\omega$ is the photon energy, $E_{x_1 x_2}$ and $E_{x_1' x_2'}$ are the energies of the final and intermediate biexciton states, respectively, and $H_{ev}^{\mathrm{intra}}$ is the part of the electron-photon coupling operator responsible for *intraband* transitions. The contribution to rate $W_2$ from *interband* transitions is cancelled by the analogous term with the reversed order of the operators $H_{xx}$ and $H_{ev}^{inter}$. In the case of *intraband* transitions, the latter term vanishes because $H_{ev}^{intra} \left| 0 \right\rangle = 0$ (in the vacuum state, the valence band in fully occupied while the conduction band is completely empty), and hence, the resulting biexciton generation rate is *nonzero*. Further, because of the significant strength of intraband transitions in NCs [14] and a large number of intermediate and final biexciton states, the process described by Eq. (3) can be highly efficient.

To estimate the CM efficiency for the virtual-biexciton mechanism, we consider lead-salt NCs using the description of electronic states developed by Kang and Wise [15]. In this formalism, electronic states are characterized by parity ($\pi$), total angular momentum ($j$), and projection thereof ($m$); $j$ is the sum of the orbital angular momentum ($l$) and the particle spin ($s$). The parities of the $2(2l + 1)$-fold degenerate energy levels are $\pi = -(-1)^l$ in the conduction band and $\pi = (-1)^l$ in the valence band. The multi-valley structure of lead salts leads to an additional four-fold degeneracy of each of the above electronic states (the inter-valley coupling is small compared to transition linewidths, and therefore, is not expected to appreciably split this degeneracy).

The selection rules for dipole-allowed intraband transitions are $\Delta j = 0, \pm 1$, $\Delta m = 0, \pm 1$, and $\pi_1 \pi_2 = -1$ [15]. Because of the requirement $\pi_1 \pi_2 = -1$, the operator $H_{ev}^{\text{intra}}$ is odd. Since the Coulomb energy $U_C(|\mathbf{r}_1 - \mathbf{r}_2|)$ is not modified under spatial inversion ($\mathbf{r}_1 \rightarrow -\mathbf{r}_1$ and $\mathbf{r}_2 \rightarrow -\mathbf{r}_2$), the matrix element $U_{b_1 b_2}^{a_1 a_2}$ is nonzero only for states that satisfy the condition $\pi_{c_1} \pi_{c_2} \pi_{d_1} \pi_{d_2} = +1$, which indicates that the operator $H_{xx}$ is even. Finally, because the vacuum state of a NC is even, the final biexciton state $|x_1 x_2\rangle$ generated by the product of operators $H_{xx}$ and $H_{ev}^{\text{intra}}$ is odd, and hence, the product of the parities of all four carriers in the final biexciton state is $-1$.

Based on parity considerations, the energy onset for CM ($\hbar \omega_{\text{CM}}$) corresponds to the lowest-energy odd biexciton state. Such a state comprises three 1S carriers ($l = 0$) and one 1P carrier ($l = 1$) and, hence, $\hbar \omega_{\text{CM}} = 3E(1S) + E(1P) = 2E_g + [E(1P) - E(1S)]$. Using the energies computed in the supporting information of Ref. [5], we obtain $\hbar \omega_{\text{CM}} = 2.25 E_g$. This result indicates that, because of the selection rules, the CM threshold is higher than the $2E_g$ limit defined by energy conservation.

To estimate the rate $W_2$, we replace the energy of the intermediate biexciton states with $2E_g$ and approximate matrix elements of operators $H_{xx}$ and $H_{ev}^{\text{intra}}$ with their average values ($U_{xx}$ and $V_{\text{intra}}$, respectively). By further performing the summation in Eq. 3 over the intermediate and the final biexciton states we obtain $W_2 \approx (2\pi / \hbar) |g'_{xx} V_{\text{intra}} U_{xx} / E_{xx}|^2 g_{xx}$, where $g'_{xx}$ is number of the intermediate biexciton resonances, while $g_{xx}$ is number of the final biexciton states with energy $E_{xx} = \hbar \omega$. We quantify the CM efficiency in terms of the ratio of the biexciton generation rate ($W_2$) and the total photogeneration rate ($W$): $f = W_2 / W$. The rate $W$ can be calculated using first-order perturbation theory: $W \approx (2\pi/\hbar)|V_{\text{inter}}|^2 g_x$, where $g_x$ is the number of single exciton resonances at energy $\hbar\omega$ and $V_{\text{inter}}$ is the average matrix element of the interband transition. Finally, we obtain $f \approx (g'_{xx})^2 (g_{xx}/g_x)(V_{\text{intra}}/V_{\text{inter}})^2 (U_{xx}/E_{xx})^2$.

We expect that the value of $U_{xx}$ is of the same order of magnitude as the e-h exchange interaction. According to Ref. [15], the latter is ca. 1 meV in PbSe NCs. Further, assuming $E_g$ of $\sim$1 eV, we obtain that $U_{xx}/E_{xx}$ is on the order of $10^{-3}$. Despite a small value of $U_{xx}/E_{xx}$, $f$ can still be a significant fraction of unity because of the factors associated with the matrix elements of optical transitions ($V_{\text{intra}}/V_{\text{inter}}$) and the spectral densities of biexcition and single-exciton states [$(g'_{xx})^2 (g_{xx}/g_x)$]. Specifically, the ratio between the strengths of intra- and interband transitions in NCs scales as $(R/a_0)^2$ ($R$ is the NCs radius and $a_0$ is the lattice constant) and, therefore, can be much greater than unity, particularly for NCs of large sizes (the $R^2$ scaling holds until NCs can still be considered as 0D systems, i.e., $R^{-1}$ is comparable or greater than the change in the $k$-vector during the intraband transition).

Further increase in $f$ occurs because of the contribution from multiple intermediate biexciton states. To estimate the number of these states, $g'_{xx}$, we take into account that they are generated from the vacuum state by the even operator $H_{xx}$, and therefore, are also even. Acting on the intermediate $|x'_1 x'_2\rangle$ state, the operator $H_{ev}^{\text{intra}}$ modifies the quantum state of only one of the four carriers, and therefore, $g'_{xx}$ is the sum of the numbers of possible intermediate states for each of these carriers. For a final S state there are two P-type intermediate states, while for a state with $l > 0$, the number of intermediate states is four. Thus, $g'_{xx} = 10, 12, 14,$ or 16 for a final state that contains three, two, one, or no S-type carriers, respectively.

Another important factor that increases the value of $f$ is a large number of final biexciton states that can greatly exceed the number of single-exciton states of the same energy. For example, the lowest-energy odd biexciton allowed in the CM process can be in one of two configurations: |1S(e)1P(h);1S(e)1S(h))> or |1P(e)1S(h);1S(e)1S(h)> (1S1S1S1P biexciton). If we account for the $2(2l + 1)$-fold degeneracy of electronic states in a single valley and the four-valley character of energy bands, the total number of degenerate biexciton states at the CM threshold is computed to be 1344. A single-exciton state that is near resonance with the 1S1S1S1P biexciton has the |1F(e)1F(h)> configuration, which has a degeneracy of 112. Therefore, already at the CM threshold, $g_{xx}$ exceeds $g_x$ by more than a factor of 10. The $g_{xx} / g_x$ ratio rapidly increases with energy [approximately as $(E - 2E_g)^{3.7}$] as illustrated in Fig. 2(b) for biexcitons with the configuration |1$l_1$(e)1$l_2$(h);1$l_3$(e)1$l_4$(h)> (1$l$ biexcitons). The degeneracy of these biexcitons is particularly high [scales with $E$ as $(E - 2E_g)^4$; Fig. 2(b)] because the electron and hole 1$l$ states are nearly equidistant [5, 15] and, further, because the spacing between these states is nearly identical in both the valence and the conduction band. By taking into account various sources of degeneracy, we calculate that $g_{xx} / g_x$ is ~250, ~2000, and ~$10^4$ for the second ($E_{xx} = 2.75E_g$), the third (3.25$E_g$), and the fourth 1$l$ biexciton state allowed in the CM process, respectively. Using these values, we further obtain that $f$ is on the order of a fraction of a percent ($f = 0.002$) at the CM threshold (estimated for $R = 3$ nm). However, because of the rapid increase in the $g_{xx}/g_x$ ratio with energy, $f$ is 0.08 for $E_{xx} = 2.75E_g$ and ~0.8 for $E_{xx} = 3.25E_g$. The overall spectral dependence of calculated values is consistent with CM efficiencies observed experimentally for PbSe NCs (inset in Fig. 2).

Since Coulomb carrier-carrier interactions do not increase overall optical absorption [10], the value of $f$ cannot exceed unity, which defines the applicability limit of the perturbation approach used here. While not affecting the total (spectrally integrated) absorption, Coulomb coupling can lead to spectral re-distribution of the oscillator strength. However, since Coulomb energies are expected to be small compared to transition linewidths, the spectral changes associated with CM are likely not significant and not readily detectable experimentally. For example, experimentally measured absorption spectra of PbSe NCs of different sizes (i.e., different band gaps) closely match each other at high spectral energies despite the difference in the CM thresholds (Fig. 3; positions of $\hbar\omega_{CM}$ are shown by arrows) and, hence, independent of whether photon absorption produces single excitons or multiexcitons.

In conclusion, we analyze a novel physical mechanism for direct generation of biexcitons in NCs, which involves excitation of a virtual biexciton from vacuum by the Coulomb interaction between two valence-band electrons, followed by its conversion into a real biexciton induced by absorption of a photon on the intraband optical transition. This mechanism is not active in bulk materials because of momentum conservation, which suppresses intraband absorption. On the other hand, because of relaxation of momentum conservation, the proposed mechanism becomes highly efficient in 0D structures. Based on our calculations it can provide a significant contribution to CM in NC materials.

This work was supported by the Chemical Sciences, Biosciences, and Geosciences Division of the Office of Basic Energy Sciences, Office of Science, U.S. Department of Energy.

## REFERENCES

*Electronic address: valery_rupasov@hotmail.com

†Electronic address: klimov@lanl.gov.

# FIGURE CAPTIONS

**Figure 1.** Direct photogeneration of a biexciton from vacuum, $|0\rangle$, via intermediate single-exciton (on the left) or intermediate biexciton (on the right) states. The final biexciton state, $|x_1 x_2\rangle$, is in resonance with the absorbed photon, while intermediate (virtual) single-exciton, $|x'\rangle$, and biexciton, $|x'_1 x'_2\rangle$, states can be off resonance (energy conservation is not required for virtual processes). The order of processes involving electron-photon and carrier-carrier Coulomb interactions for the "virtual-biexciton" channel is reversed with respect to that for the "virtual-exciton" channel. $H_{ll}$ and $H_{xx}$ are the components of the Coulomb-interaction operator responsible for impact-ionization and biexciton generation, respectively. $H_{ev}^{\text{inter}}$ and $H_{ev}^{\text{intra}}$ are the operators of the electron-photon interaction responsible for the interband and intraband transitions, respectively.

**Figure 2.** (a) Spectral dependence of quantum efficiencies measured for bulk PbS (data from Ref. 13) as well as PbS NCs and PbSe NCs (from Ref. 6). (b) Calculated degeneracy factors for four low-energy $1l$ biexcitons (solid circles) and corresponding near-resonant single exciton states (open squares) along with their ratio (solid diamonds) shown as a function of energy in access of $2E_g$. Solid lines are fits to $(E - 2E_g)^m$, with $m = 4$ ($g_{xx}$) and 3.7 ($g_{xx}/g_x$); the dotted line is a guide for eye. (c) Comparison of calculated (solid circles) and measured (open squares) values of $f = W_2/W$.

**Figure 3.** Spectral dependence of the absorption coefficient ($\alpha_0$) of PbSe NCs of three different mean radii (indicated in the figure). Arrows show the onset for CM calculated from $\hbar\omega_{\text{CM}} = 2.85\, E_g$.

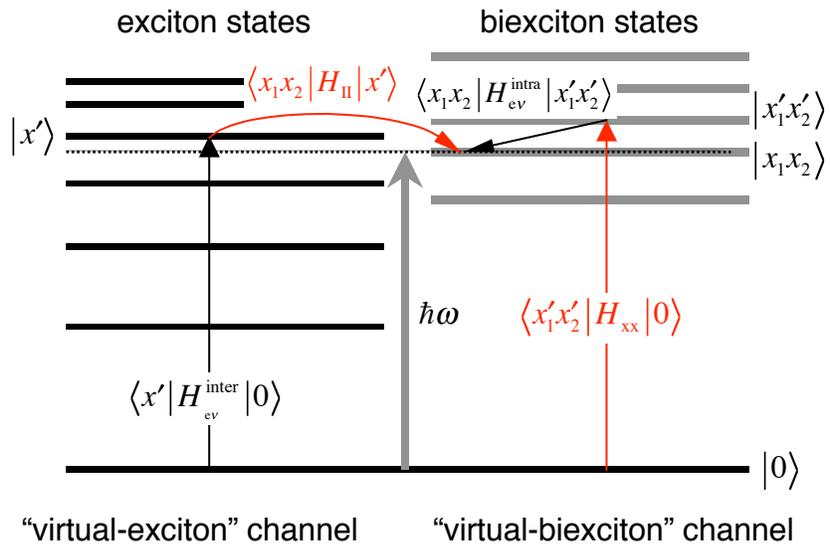

Figure 1.

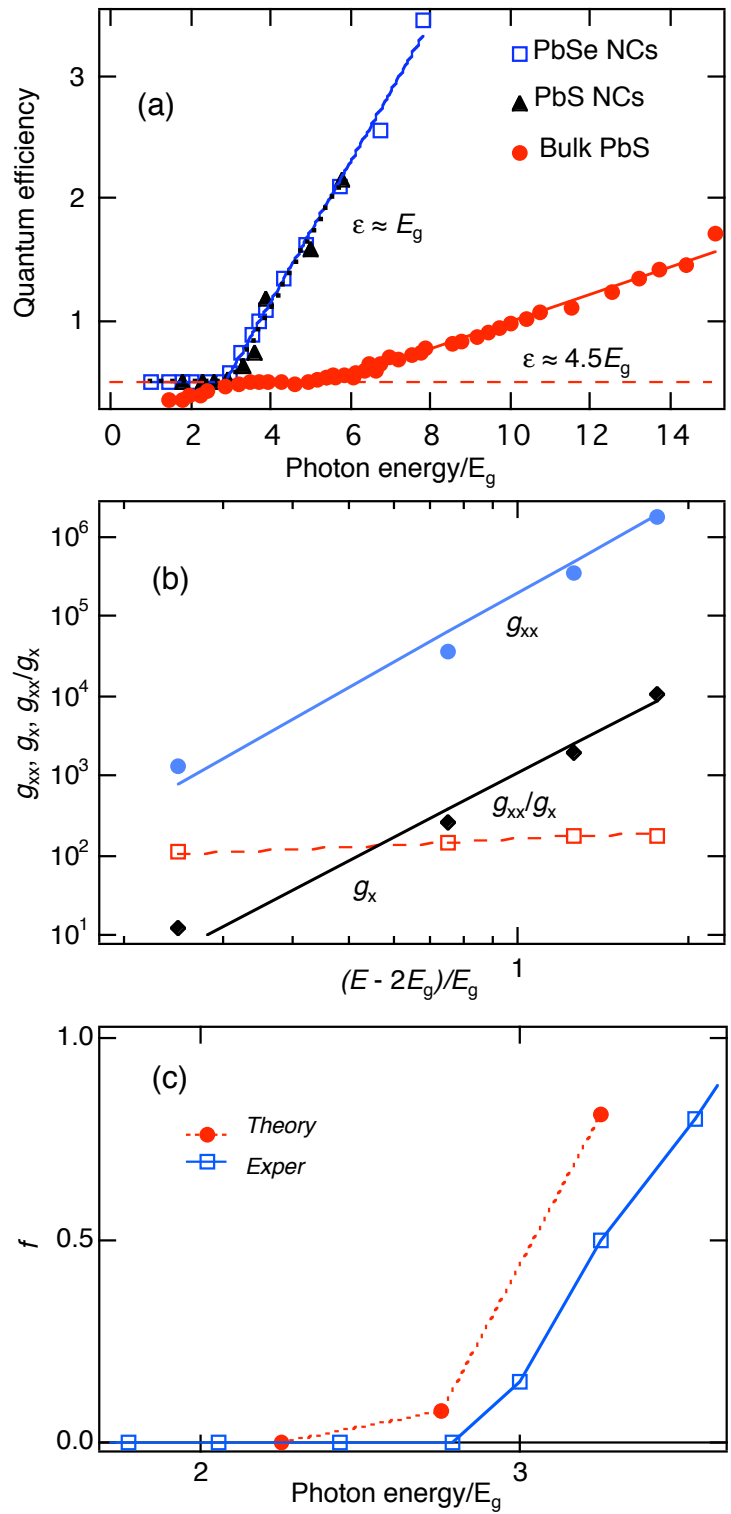

Figure 2.

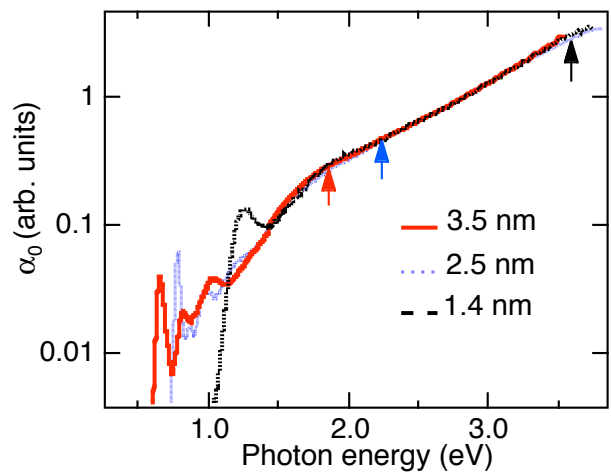

Figure 3.